\documentclass[conference]{IEEEtran}
\IEEEoverridecommandlockouts

\usepackage{bm}
\usepackage{multirow}
\usepackage{multicol}
\usepackage{makecell}
\usepackage{booktabs}

\usepackage{cite}
\usepackage{array}
\usepackage{amsmath,amssymb,amsfonts}
\usepackage{algorithmic}
\usepackage{graphicx}
\usepackage{textcomp}
\usepackage{xcolor}

\begin{document}

\title{Automatic Inspection Based on Switch Sounds of \\
Electric Point Machines}

\author{\IEEEauthorblockN{
Ayano Shibata$^{\star}$,
Toshiki Gunji$^{\star}$,
Mitsuaki Tsuda$^{\star}$,\\
Takashi Endo$^{\dagger}$,
Kota Dohi$^{\dagger}$,
Tomoya Nishida$^{\dagger}$, and
Satoko Nomoto$^{\dagger}$}
\IEEEauthorblockA{$^{\star}$\textit{East Japan Railway Co.} \quad $^{\dagger}$\textit{Hitachi, Ltd.}
}}

\maketitle

\begin{abstract}
Since 2018, East Japan Railway Company and Hitachi, Ltd. have been working to replace human inspections with IoT-based monitoring. 
The purpose is Labor-saving required for equipment inspections and provide appropriate preventive maintenance. 
As an alternative to visual inspection, it has been difficult to substitute electrical characteristic monitoring, and the introduction of new high-performance sensors has been costly. 
In 2019, we implemented cameras and microphones in an ``NS'' electric point machines to reduce downtime from equipment failures, allowing for remote monitoring of lock-piece conditions. 
This method for detecting turnout switching errors based on sound information was proposed, and the expected test results were obtained. 
The proposed method will make it possible to detect equipment failures in real time, thereby reducing the need for visual inspections.
This paper presents the results of our technical studies aimed at automating the inspection of electronic point machines using sound, specifically focusing on ``switch sound'' beginning in 2019.
\end{abstract}

\section{Introduction}
\label{sec:intro}
In the operation and maintenance of railway signaling systems, inspections of signaling equipment are currently conducted on-site by personnel at regular intervals. 
In the event of equipment failure, corrective maintenance is carried out in response. Additionally, periodic replacement of equipment has been performed based on the replacement cycles specified by the equipment manufacturers.

With the advancement of IoT technologies, there is a growing movement toward optimizing operation and maintenance through the adoption of Condition-Based Maintenance (CBM). 
This approach aims to enhance inspection efficiency through data analysis, implement preventive maintenance, and manage equipment replacement based on actual equipment conditions. 
As part of this initiative, studies have also been conducted on electronic point machines, which serve as train route control devices.

We began efforts to reduce inspections by monitoring the sound, focusing on ``switch sound,'' which are widely used in machine maintenance. 
In this article, we will introduce JR East's efforts to replace on-site inspections by humans by focusing on the switch sound of electric point machines and grasping the condition of electric point machines.

\begin{figure}[th]
    \center
    \includegraphics[width=1.0\linewidth,clip]{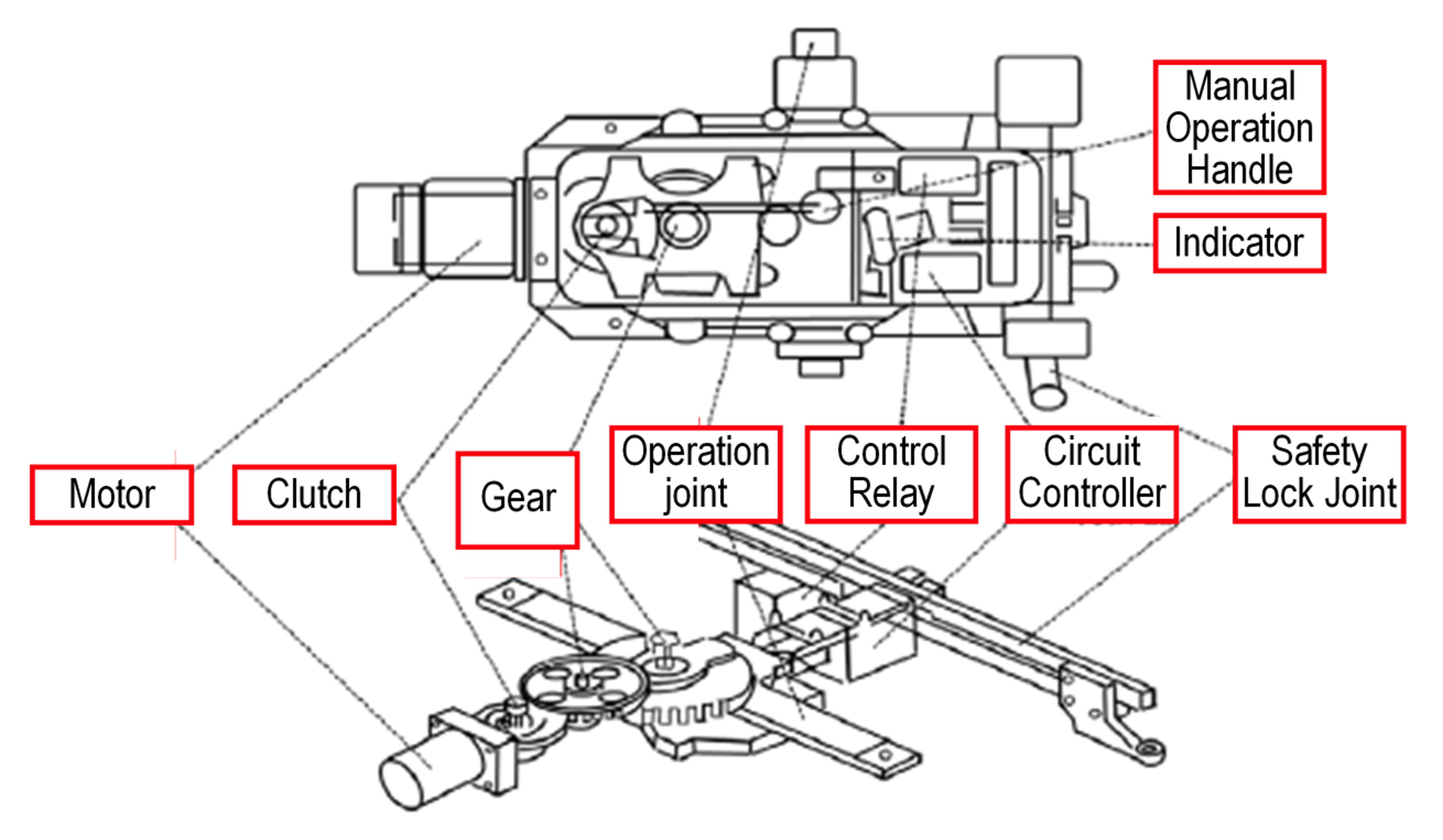}
    \caption{Electric Motor Point Machine Overview}
    \label{fig:electric}
\end{figure}

This paper focuses on NS-type point machines, which are equipment installed by JR East. This is an overview of an NS-type point machine (Figure \ref{fig:electric}).

\section{Development of ``Switching Phase'' Detection Methodology}
\label{sec:switch}

This section introduces how to use sound to identify units of time during which a motor is running that can be inspected and replaced (hereafter referred to as ``Switching Phase''). 
Conventionally, the only switching phases that could be identified solely from the input and output of the interlocking device, i.e., the electric railway point machine, were the issuance of a switching start command and the reception of a switching completion response. 
For this reason, in order to obtain new switching phase information, it was common to either add a new high-performance sensor or estimate it based on the time of each phase observed in the past. 
Therefore, we proposed and verified a method to recognize switching phases based solely on sound information.

\subsection{Separating Transformation Phases and Selecting Targets for Each ``Switching Phase''}
\begin{figure*}[t]
    \center
    \includegraphics[width=0.9\linewidth,clip]{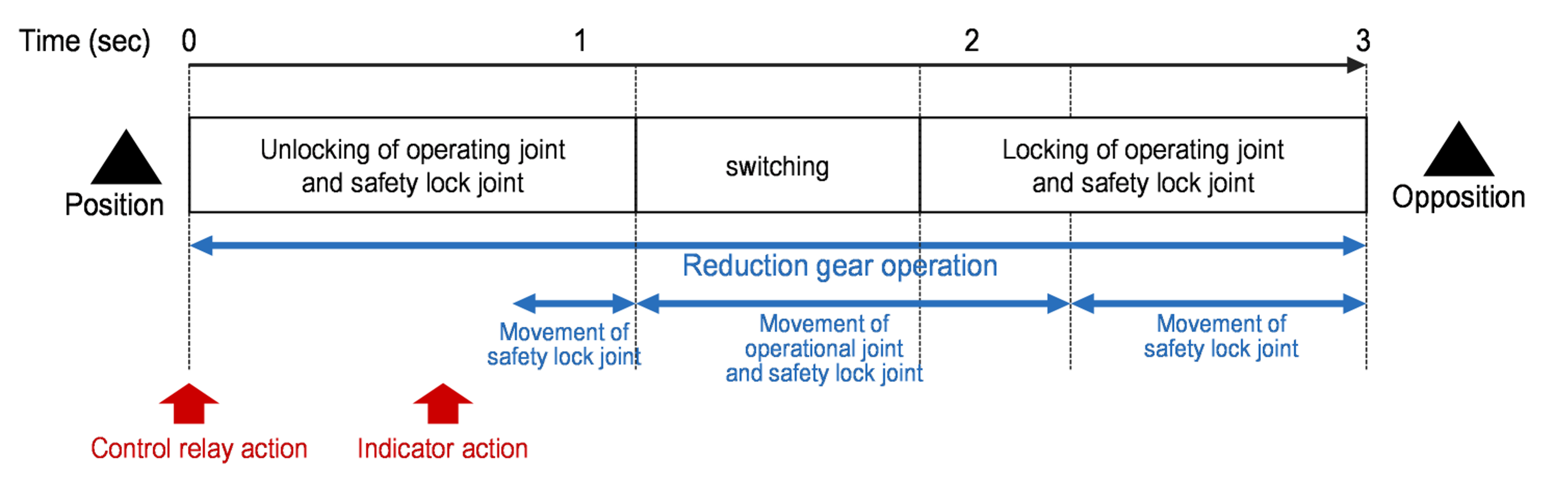}
    \caption{The Seven Phases of Electric Motor Point Moving}
    \label{fig:seven}
\end{figure*}

The ``NS'' electric point machines have seven “Switching Phases” in that switching state (Figure \ref{fig:seven}).
By linking these switching operations and switching sounds, we were able to classify the characteristics of the malfunctions we wanted to capture (Table \ref{tab:classification}).
\begin{table*}[t]
\centering
\caption{Classification of Abnormalities that Can be Detected by Each Phase}
\label{tab:classification}
{\scriptsize
\begin{tabular}{llcccccccccc}
\toprule
&&  
\multirow{3}{*}{\makecell[c]{Grease\\Inside Dev.}} & 
\multirow{3}{*}{\makecell[c]{Contact}} & 
\multirow{3}{*}{\makecell[c]{Lock}} & 
\multirow{3}{*}{\makecell[c]{Gear}} &
\multirow{3}{*}{\makecell[c]{Motor}} &
\multirow{3}{*}{\makecell[c]{Collision\\with\\Inclusions}} &
\multirow{3}{*}{\makecell[c]{Collision\\with Snow\\Melting Dev.}} &
\multirow{3}{*}{\makecell[c]{Grease\\Outside Dev.}} &
\multirow{3}{*}{\makecell[c]{Torque\\Fluctuation}} &
\multirow{3}{*}{\makecell[c]{Insufficient\\Control\\Voltage}}\\
\multicolumn{2}{l}{Type of Malfunctions}\\\\
\midrule
\multirow{7}{*}{\rotatebox[origin=c]{90}{Phase}}
 & Starting Routine & --- & --- & --- & --- & --- & --- & --- & --- & --- & \checkmark \\
 & Idle before Moving & \checkmark & --- & --- & \checkmark & \checkmark & --- & --- & --- & --- & --- \\
 & Deactivate Safety & \checkmark & --- & \checkmark & \checkmark & \checkmark & --- & --- & --- & --- & --- \\
 & Moving Rail & --- & \checkmark & --- & \checkmark & \checkmark & \checkmark & \checkmark & \checkmark & \checkmark & --- \\
 & Activate Safety & \checkmark & --- & \checkmark & \checkmark & \checkmark & --- & --- & --- & --- & --- \\
 & Idle after Moving & \checkmark & --- & --- & \checkmark & \checkmark & --- & --- & --- & --- & --- \\
 & Ending Routine & --- & --- & --- & --- & --- & --- & --- & --- & --- & --- \\
\midrule
\multirow{5}{*}{\rotatebox[origin=c]{90}{Inspection}} & 
%\multirow{5}{*}{\makecell[l]{Inspection}} & 
\multirow{4}{*}{\makecell[l]{Action}} & 
\multirow{4}{*}{\makecell[c]{Grease\\Injection to\\Inside Dev.}} & 
\multirow{4}{*}{\makecell[c]{Contact\\Force}} & 
\multirow{4}{*}{\makecell[c]{Gap of\\Locking\\Piece}} & 
\multirow{4}{*}{\makecell[c]{Check\\Status}} & 
\multirow{4}{*}{\makecell[c]{Check\\Status}} &
\multirow{4}{*}{\makecell[c]{Check\\Status}} &
\multirow{4}{*}{\makecell[c]{Dist. Point\\M and Snow\\Melting Dev.}} &
\multirow{4}{*}{\makecell[c]{Grease\\Injection to\\Outside Dev.}} & 
\multirow{4}{*}{\makecell[c]{Cleaning\\and\\Inspection}} & 
\multirow{4}{*}{\makecell[c]{---}} \\\\\\\\
 & Cost Affective & Very High & High & Avg. & Avg. & Avg. & Avg. & Low & Low & Very Low & --- \\
\bottomrule
\end{tabular}
}
\end{table*}

\subsection{Overview of ``Switching Phase'' Separation by Sound}
The sounds produced by the operation of each component of the point machine have distinct spectral shapes. 
We aimed to determine the sound volume of each component using semi-supervised non-negative matrix factorization (SNMF)~\cite{lee2010}. 
By labeling the time ranges of each component's operation within the spectrograms of the operational sounds, including switching actions, we performed non-negative factorization to obtain matrices corresponding to each component's timbre. 
Using these matrices, we conducted non-negative matrix factorization on the spectrogram of the point machine’s operational sounds to estimate the volume (activation) of each sound. 
Based on these results, we attempted to estimate the operational phases.

\subsection{Results}
\begin{figure}[th]
    \center
    \includegraphics[width=1.0\linewidth,clip]{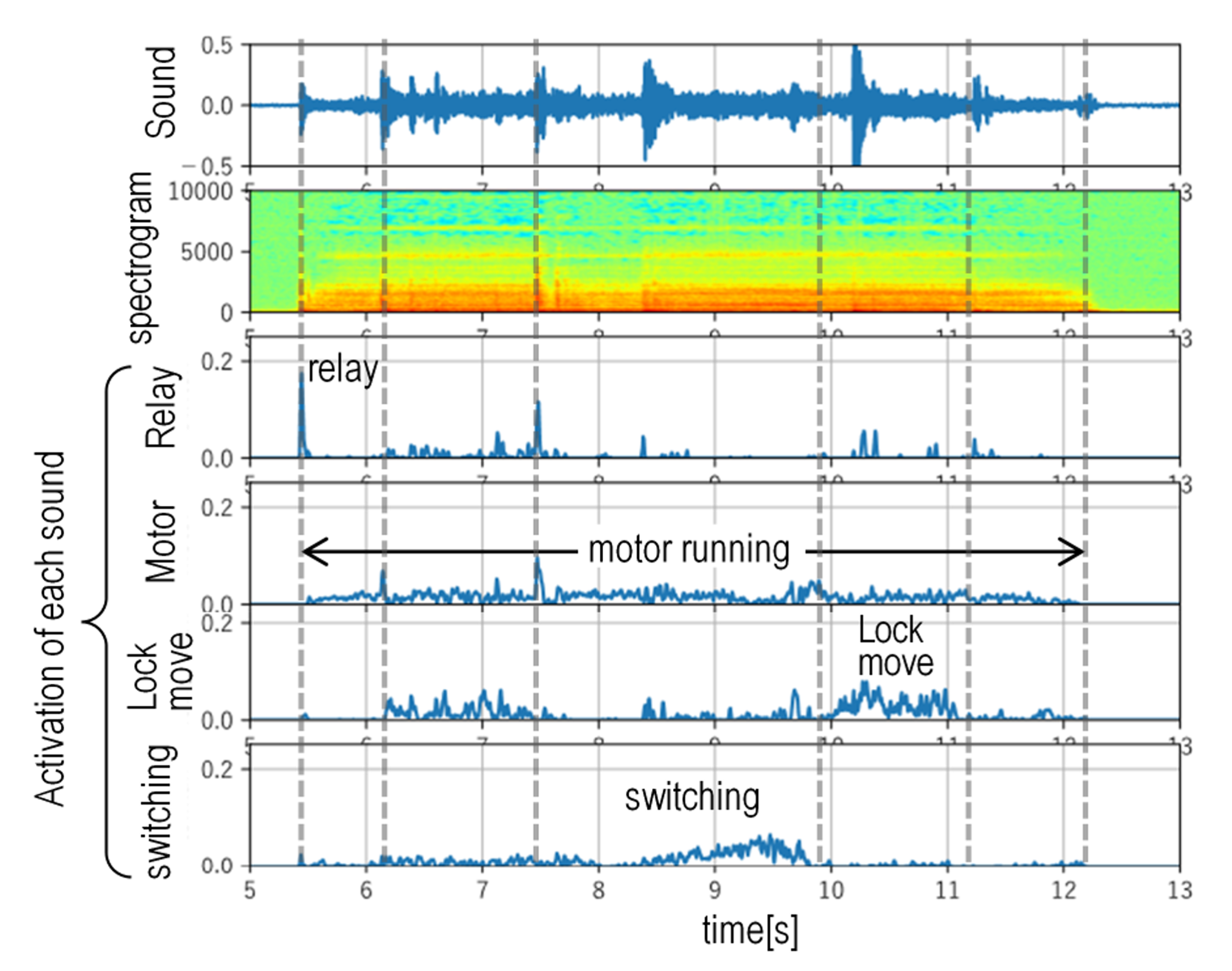}
    \caption{Activation of Operational Sounds of the Point Machine}
    \label{fig:activation}
\end{figure}

The results of the correspondence between the phases of the point machine and the activation of SNMF are shown in Figure \ref{fig:activation}. 
It was confirmed that the decomposition into phases is possible through the sounds originating from the relay, the motor operation, the movement of the locking piece, and the switching operation sound associated with the movement of the rod.

\begin{figure*}[t]
    \center
    \includegraphics[width=1.0\linewidth,clip]{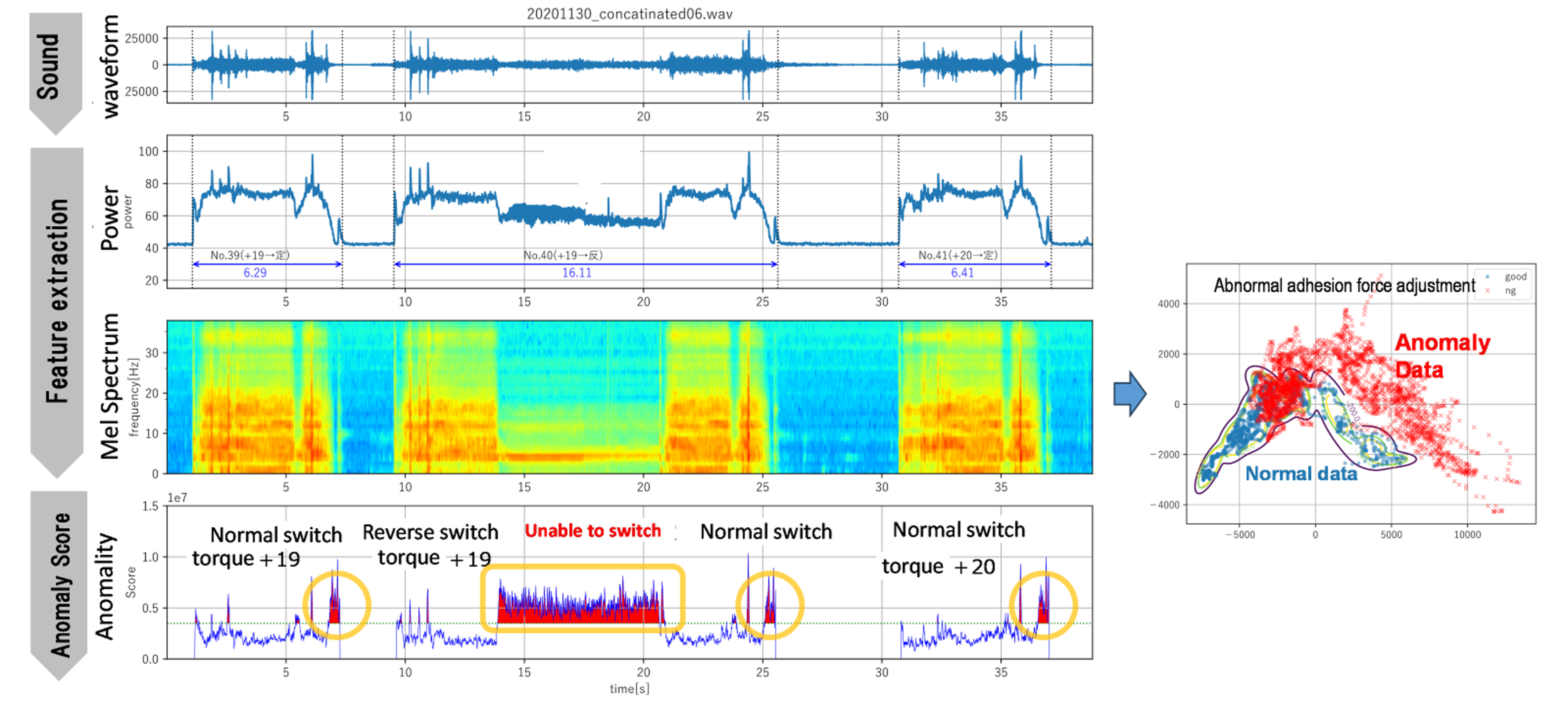}
    \caption{Analyzing Features from Sound and Visualizing the Distribution of Features}
    \label{fig:analyzing}
\end{figure*}
\begin{figure}[th]
    \center
    \includegraphics[width=1.0\linewidth,clip]{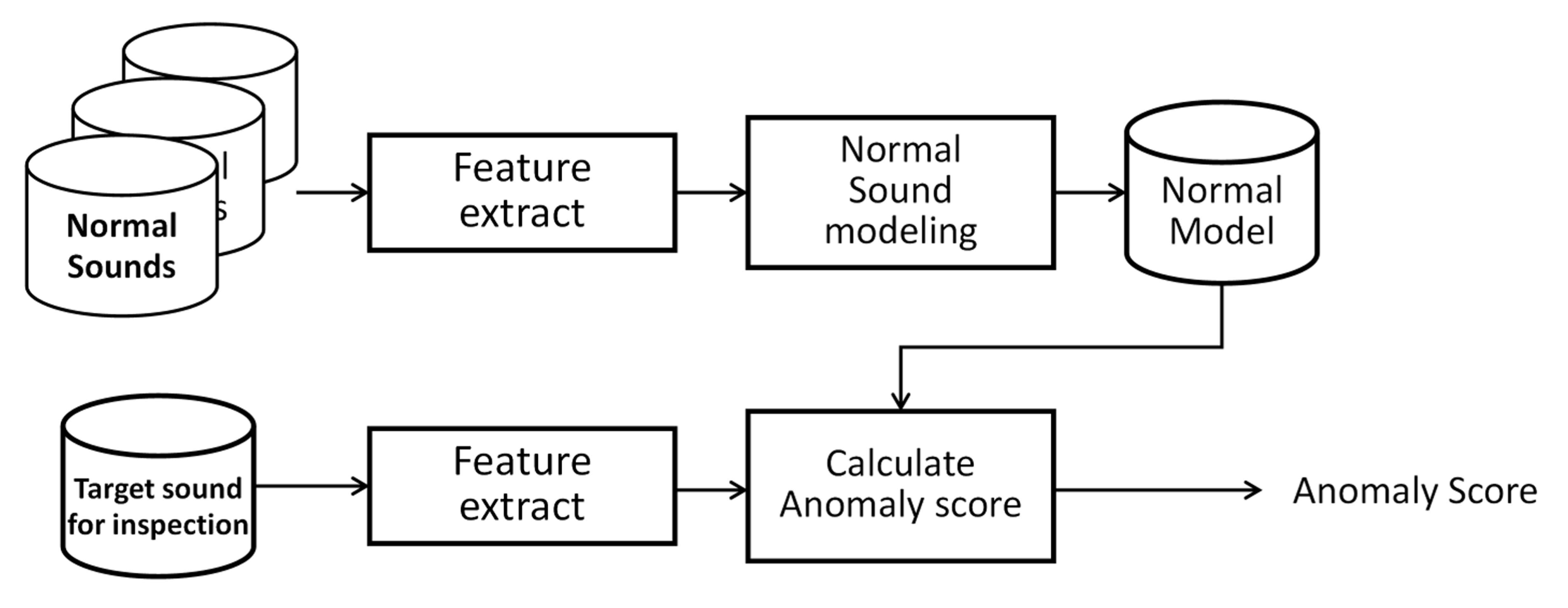}
    \caption{Block Diagram of Acoustic Diagnosis}
    \label{fig:block}
\end{figure}

\section{Development of an Inspection Model}
\label{sec:inspection}
Next, we developed a program to calculate the degree of abnormality in the switch sound with aim of reducing inspection work. 
As being cost-effective, we selected the three types that significantly impact inspection reduction: ``grease deterioration inside the point machine,'' ``abnormal lock deviation,'' and ``abnormal adhesion force adjustment''.
We compared the sounds in normal and abnormal conditions and analyzed the distribution of feature values.
The program quantified the degree of abnormality for each adjustment condition, visualized these conditions by graphing, and confirmed that setting a threshold value allows for effective abnormality detection.

\subsection{Overview of Anomaly Detection by Sound}
Based on the switch sound ``Switching Phase'' separation completed, we compared the sounds in normal and abnormal states for ``grease deterioration inside the point machine,'' ``abnormal lock deviation,'' and ``abnormal adhesion force adjustment,'' which have a high effect on reducing inspections, and analyzed the distribution of feature values (Figures \ref{fig:analyzing} and \ref{fig:block}).

\subsection{Proposal of an Inspection Model}
\begin{figure*}[t]
    \center
    \includegraphics[width=0.8\linewidth,clip]{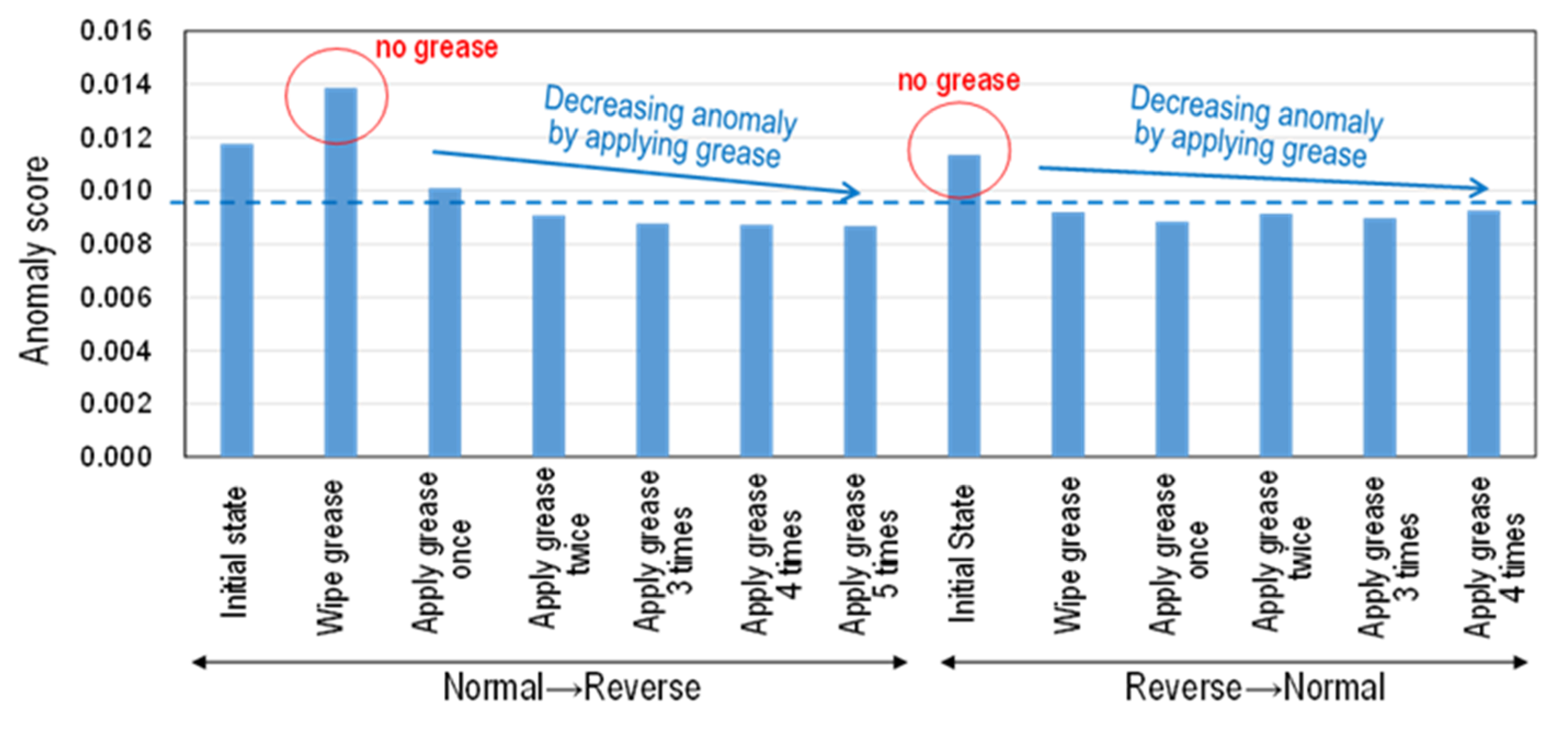}
    \caption{Results of Detection of Grease Deterioration in Electric Point Machines}
    \label{fig:grease}
\end{figure*}

\begin{figure*}[t]
    \center
    \includegraphics[width=0.5\linewidth,clip]{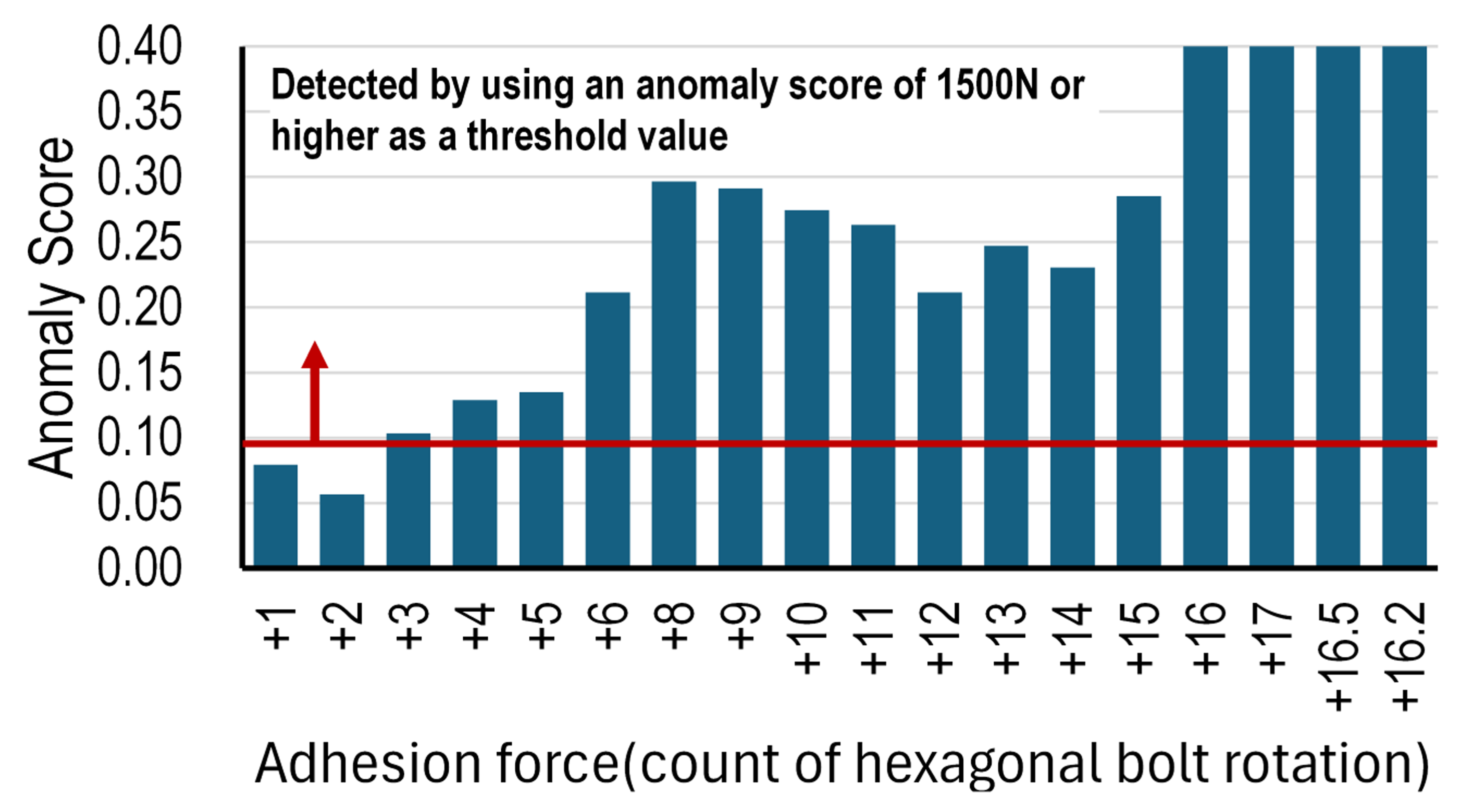}
    \caption{Results of Anomaly Detection in the Adhesion Force of Electric Point Machines}
    \label{fig:adhesion}
\end{figure*}

\begin{figure*}[t]
    \center
    \includegraphics[width=0.8\linewidth,clip]{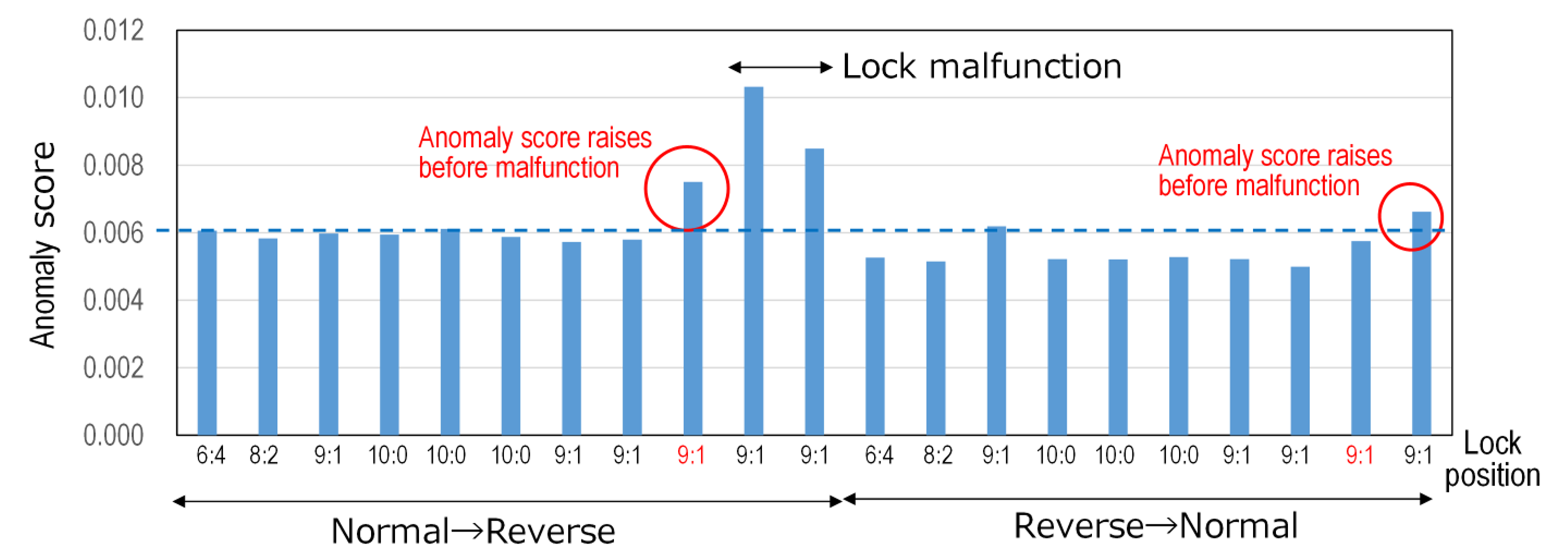}
    \caption{Results of Anomaly Detection for Locking Piece Position of Electric Point Machines}
    \label{fig:locking}
\end{figure*}

Subsequently, based on the abnormal sound detection model, we developed a simple program to calculate the degree of abnormality of the switch sound. 
This time, judgment was made on three items: grease deterioration, lock abnormality, and adhesion abnormality. 
The program quantified the degree of abnormality for each adjustment condition, visualized the condition by graphing, and verified that abnormality detection is possible by setting a threshold value (Figures \ref{fig:grease}, \ref{fig:adhesion}, and \ref{fig:locking}).
\ref{fig:grease}.

\subsubsection{Grease Deterioration inside the Point Machine}
In this section the grease in the point machine is maintained by periodic lubrication; however, the consumption of grease varies depending on operational conditions, and currently, it is not possible to ascertain the grease consumption status without physically visiting the site.
Consequently, lubrication is often performed earlier than necessary to allow for ample time. 
If remote automatic monitoring of the grease quantity condition can be achieved, it would enable CBM. 

In this experiment, all the grease was initially wiped off with a cloth, and a relationship between the operational sounds recorded by a microphone and the amount of lubrication applied was evaluated. 
This approach aims to investigate the extent to which the reduction of grease can be inferred from the acoustic data. Normal model is implemented by VIDNN algorithm~\cite{suefusa2020}. 
Once grease is wiped away from point machine, we treated the point machine as anomaly status. And then apply grease several times using grease injector. After 4 times of grease injection, we treated the point machine as normal grease state.

The initial state refers to the operational sound during switching conducted before the experiment. 
After completely wiping off the grease from the point machine in the initial state, lubrication was performed by pushing the grease injector one push at a time, and switching operations were carried out for each state of the grease. 
When the grease was completely wiped off, the deviation from the normal sound became greater, and the degree of abnormality score increased to a point where it is distinguishable from the normal state as shown in Figure \ref{fig:grease}. 

\subsubsection{Abnormal Adhesion Force Adjustment}
If there is a misadjustment in the adhesion force at the tip of the tongue rail of the switch, a failure to switch may occur. 
In this study, we rotated the hexagonal bolts that adjust the adhesion force of the tongue rail, measuring the sound during the switching operation while varying the adhesion force. 
We conducted experiments to determine whether the acoustic detection is possible when a strong adhesion force is applied.

In the experiment, the adhesion force of the tongue rail started from a normal state, and the hexagonal bolts were rotated by $1/6$ of a turn to vary the adhesion force until a switching failure occurred. 
We conducted switching experiments to evaluate whether the acoustic signals could detect abnormalities in the adhesion force. 
Additionally, if detection was possible, we assessed how far in advance the detection could occur before the switching anomaly manifested.

Figure \ref{fig:adhesion} shows the results of adjusting the adhesion level of the tongue rail by rotating the hexagonal bolts in increments of $1/6$ turn. 
As the adhesion level was increased, the operational sound during the switching process was recorded, and the power and spectrum were analyzed. 
Changes in power and spectrum were observed around the point where the hexagonal bolts were rotated by $6/6$ turns (more than $+6$ of adhesion force), along with variations in the switching time. 
A switching failure occurred after the hexagonal bolts were rotated by $17/6$ turns.

\subsubsection{Abnormal Lock Deviation}
The locking piece of an point machine moves in conjunction with the locking rod as the tongue rail operates. When it reaches either the normal or reverse position, the locking piece is inserted into the notch of the locking rod to prevent the tongue rail from moving. 
There is a gap between the locking piece and the locking rod, and the amount of this gap must be the same on both sides. 
The ratio of the left and right gaps is expressed as a proportion, with a $5:5$ ratio indicating a normal state. If the ratio deviates, it may become $0:10$ or $10:0$. As the deviation increases, the locking piece may fail to enter the notch of the locking rod, which is considered one of the causes of switching failures \cite{ryuo2015}.
This chapter attempts to detect the misalignment of the locking piece using acoustic signals.

Starting from the locked position adjusted to a $5:5$ ratio, the locking position was gradually shifted towards a $10:0$ ratio while recording the sound until a failure to switch occurred. 
After each switching operation, the locking position was checked, which is why the positions of $10:0$ and $9:1$ are noted. 
The adjustment of the locking position is made by modifying the position of the locking rod to increase the balance of the gaps on both sides.

In the transition from normal to reverse position, the anomaly score increased significantly from the $9:1$ ratio (third instance) before the occurrence of a locking failure as shown in Figure \ref{fig:locking}. 
Therefore, it was confirmed that anomalies could be detected even before the locking failure manifested.

\subsection{Summary of Experiment}
Experiments for predictive detection using acoustic diagnostic technology were conducted on three types of anomalies: ``grease deterioration inside the point machine,'' ``lock adjustment anomalies,'' and ``adhesion force adjustment anomalies.'' 
In the evaluation experiments at the training center, we created a gradual deterioration from a normal state to a complete anomaly state on an actual point machine, recording the operational sounds throughout the process. 
It was confirmed that detection through sound was possible at stages prior to the occurrence of a switching failure in each case. 
From these results, it can be inferred that CBM is feasible through acoustic diagnosis of the point machine's state.

\section{Development of Noise Removal Methodology}
This section introduces the effectiveness of noise removal. The recorded sound included background noise from passing trains, rain, birds, and cars. We developed an AI model to separate the necessary switch sound from the background noise.

\subsection{Overview of Noise Removal}
\begin{figure}[t]
    \center
    \includegraphics[width=0.9\linewidth,clip]{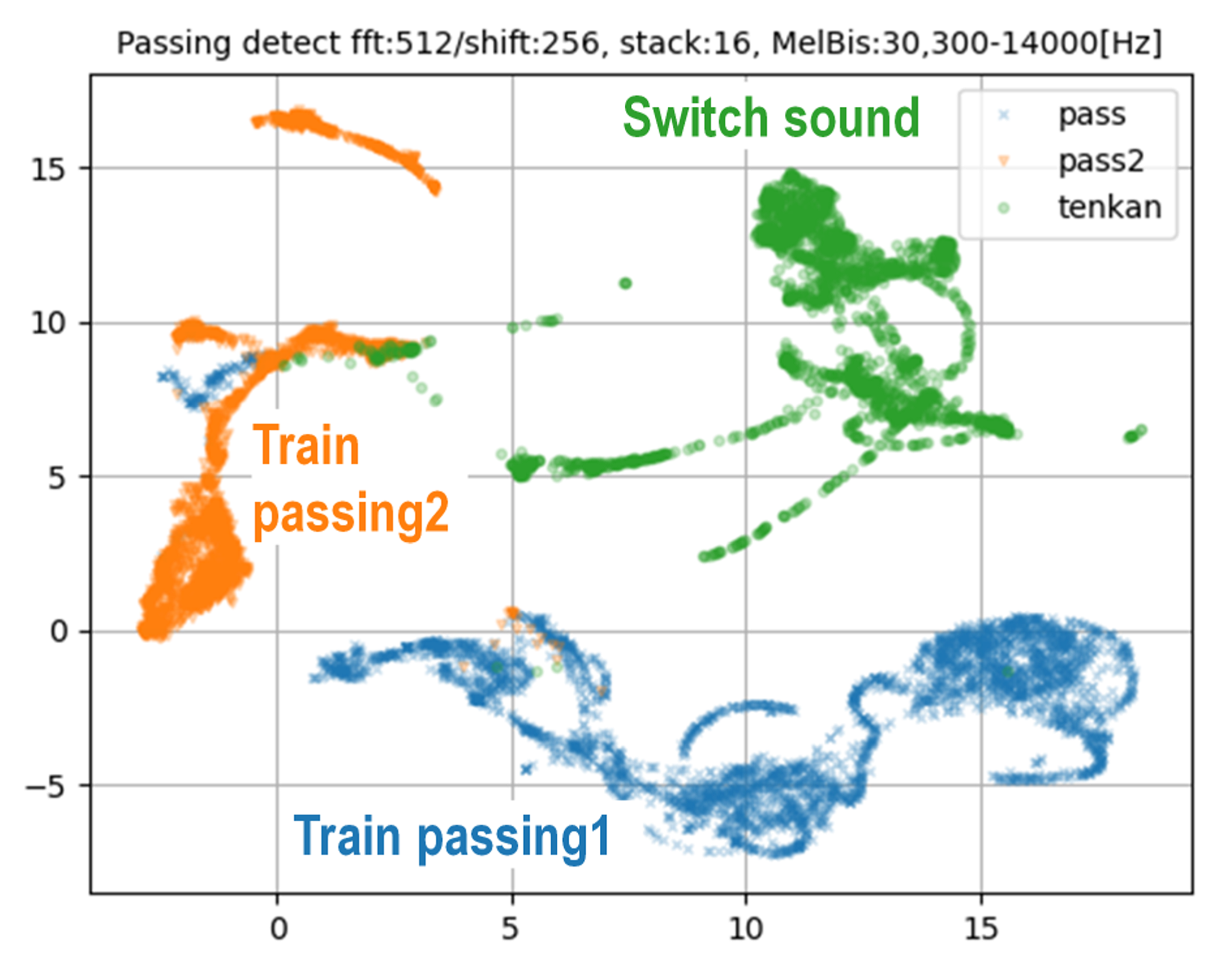}
    \caption{Distribution of Frequency Features: Switching Sounds vs. Other Sounds}
    \label{fig:distribution}
\end{figure}

The model estimation and parameter derivation were performed using the following procedure.
\begin{enumerate}
\item The audio signals include various sounds, such as the noise of trains passing on adjacent rails, the sound of Shinkansen traversing overhead viaducts, and vehicle passing noises. 
Since external noise is generally not synchronized with the switching operation, the same disturbance does not occur across multiple switching sounds. 
Therefore, operation sounds containing significant disturbances are excluded from the analysis. 
As shown in Figure \ref{fig:distribution}, switching sounds and passing noises can be distinguished within the acoustic feature space, allowing switching sounds affected by external disturbances to be excluded from the anomaly detection evaluation.
\item Noise removal is performed by excluding frequency bands that are unnecessary for diagnosing the switching operation of the turnout mechanism. 
After such noise removal, disturbances that consistently affect two or more switching operations at the same timing and cause abnormality levels are diagnosed as anomalies.
\end{enumerate}

\subsection{Results}
\begin{figure*}[th]
    \center
    \includegraphics[width=0.8\linewidth,clip]{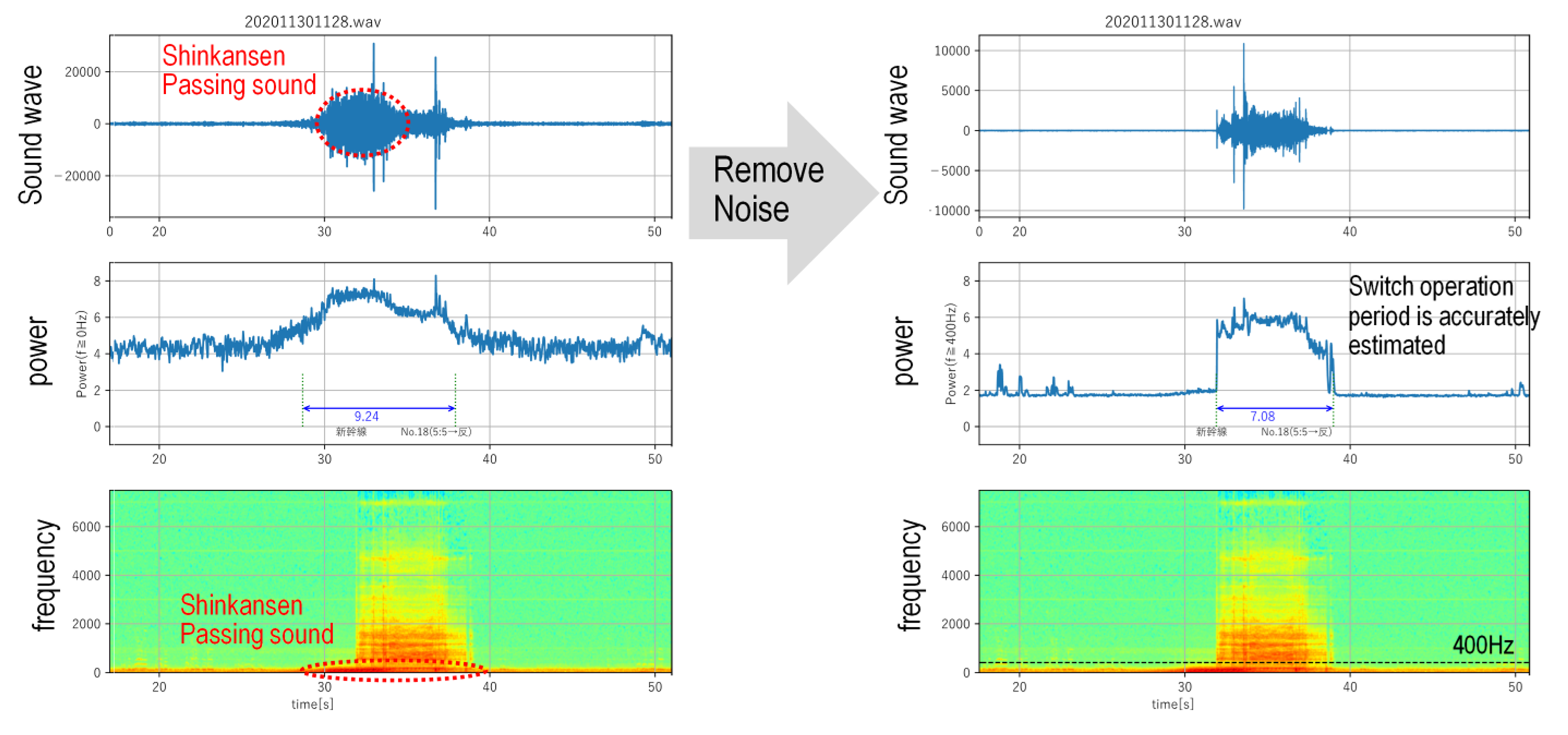}
    \caption{Noise Filtering Process}
    \label{fig:noise}
\end{figure*}

Figure \ref{fig:noise} shows the measured waveform containing Shinkansen passing sounds and the waveform after removing those sounds. 
It was confirmed that the components of the turnout switching sound remain, and the extraction of the switching period functions effectively.

\section{Conclusion}
We investigated an anomaly estimation model focusing on the switch sound of electric point machines and considered its use as an inspection substitute. 
Therefore, we verified the detection of equipment defects through AI-based voice analysis and obtained the expected results. 
The novelty of this sound-based anomaly estimation method lies in its ability to assess equipment soundness automatically through IoT device monitoring. 
This method can also optimize the replacement cycle by allowing us to develop a strategy for replacing electronic point machine equipment. 
The modeling methodology can extend to other devices, suggesting potential applications and enabling the establishment of a deterioration judgment technique.

As a future direction, it is necessary to accumulate and compare additional equipment data to validate the applicability of the model proposed in this study, with the aim of improving detection accuracy.

\section{Acknowledgment}
We would like to express our sincere gratitude to Kyosan Electric Manufacturing Co., Ltd. that manufactured locking monitor cameras for their valuable support and contributions.

\bibliographystyle{IEEEtran}
\bibliography{mybib}

\begin{thebibliography}{1}
\providecommand{\url}[1]{#1}
\def\UrlFont{\rmfamily}
\providecommand{\newblock}{\relax}
\providecommand{\bibinfo}[2]{#2}
\providecommand\BIBentrySTDinterwordspacing{\spaceskip=0pt\relax}
\providecommand\BIBentryALTinterwordstretchfactor{4}
\providecommand\BIBentryALTinterwordspacing{\spaceskip=\fontdimen2\font plus
\BIBentryALTinterwordstretchfactor\fontdimen3\font minus \fontdimen4\font\relax}
\providecommand\BIBforeignlanguage[2]{{%
\expandafter\ifx\csname l@#1\endcsname\relax
\typeout{** WARNING: IEEEtran.bst: No hyphenation pattern has been}%
\typeout{** loaded for the language `#1'. Using the pattern for}%
\typeout{** the default language instead.}%
\else
\language=\csname l@#1\endcsname
\fi
#2}}

\bibitem{lee2010}
H.~Lee, J.~Yoo, and S.~Choi, ``Semi-supervised nonnegative matrix factorization,'' \emph{IEEE Signal Processing Letters}, vol.~17, no.~1, pp. 4--7, 2010.

\bibitem{suefusa2020}
K.~Suefusa, T.~Nishida, H.~Purohit, R.~Tanabe, T.~Endo, and Y.~Kawaguchi, ``Anomalous sound detection based on interpolation deep neural network,'' in \emph{Proc. {IEEE} International Conference on Acoustics, Speech and Signal Processing (ICASSP)}, 2020, pp. 271--275.

\bibitem{ryuo2015}
S.~Ryuo and K.~Kawasaki, ``Short-term prediction of locking-error position based on locking monitor data of an electric point machine,'' \emph{IEEJ Transactions on Industry Applications}, vol. 135, no.~7, pp. 741--745, 2015.

\end{thebibliography}

\end{document}